\documentclass[twocolumns,traditabstract]{aa} % for the abstract without structuration

\usepackage{graphicx}
\usepackage{txfonts}
\usepackage{natbib}
\usepackage{longtable}
\begin{document}

\title{
Molecular shells in IRC+10216: tracing the mass  loss history
\thanks{This work was based on observations carried out with the IRAM 30-meter telescope. IRAM is 
supported by INSU/CNRS (France), MPG (Germany) and IGN (Spain).}
}
\titlerunning{Molecular Shells in IRC+10216}
\authorrunning{Cernicharo et al.}

\author{
J. Cernicharo\inst{1}, N. Marcelino\inst{2}
M. Ag\'undez\inst{1}, M. Gu\'elin\inst{3,4} }

\institute{ICMM. CSIC. Group of Molecular Astrophysics. C/ Sor Juana In\'es de la Cruz 3. Cantoblanco,
28049 Madrid. Spain; jose.cernicharo@csic.es
\and
National Radio Astronomy Observatory,
520 Edgemont Road, Charlottesville, VA 22903, USA
\and Institut de
Radioastronomie Millim\'etrique, 300 rue de la Piscine, 38406
Saint Martin d'H\`eres, France 
\and
LERMA, Observatoire de Paris, PSL Research University, CNRS, UMR 8112, F-75014, Paris France}
%,guelin@iram.fr}

\date{Received July, 7, 2014; accepted}

\abstract{

{Thermally-pulsating AGB stars provide three-fourths of the matter returned to the interstellar medium. 
The mass and chemical composition of their ejecta largely control the chemical evolution of galaxies. 
Yet, both the mass loss process and the gas chemical composition remain poorly understood.} {We present  
maps of the extended $^{12}$CO and $^{13}$CO emissions in IRC+10216, the envelope of CW~Leo, the high mass loss 
star the closest to the Sun. IRC+10216 is nearly spherical and expands radially with a velocity of 14.5 km/s. 
The observations were made On-the-Fly with the IRAM 30-m telescope;
their sensibility, calibration, and angular resolution are far higher than all previous studies.
The telescope resolution at $\lambda= 1.3$ mm (11$''$ HPBW) corresponds to an expansion time of 500 yr.}{ The CO 
emission consists of a centrally peaked pedestal and a series of bright, nearly spherical shells. 
It peaks on CW~Leo and remains relatively strong up to $r_{phot}=180''$. Further out the emission becomes very weak and 
vanishes as CO gets photodissociated. As CO is the best tracer of the gas up to $r_{phot}$, the maps show the mass 
loss history in the last 8000 yr. The bright CO shells denote over-dense regions. They show that the mass loss process 
is highly variable on timescales of hundreds of years. The new data, however, infirm
%\LEt {or do you mean "confirm"} 
previous claims of a strong decrease 
of the average mass loss in the last few thousand years. The over-dense shells are not perfectly concentric and extend 
farther to the N-NW. The typical shell
separation is $800-1000$ yr in the middle of the envelope, but seems to increase outwards. The shell-intershell brightness contrast is $\geq 3$.} {All those key features can be accounted for if CW~Leo has a companion star 
with a period $\simeq 800$ yr that increases the mass loss rate when it comes close to periastron. Higher angular resolution observations are needed to fully resolve the dense shells and measure the density contrast.
The latter plays an essential role in our understanding of the envelope chemistry.}
}


\maketitle

\section{Introduction}

The mass and chemical composition of stellar ejecta are keys to our
understanding of Galactic chemical evolution. The main source of
ejecta are Supernova explosions and AGB star winds. The latter, in
particular, the winds of thermally pulsing AGB (TP-AGB) stars,
provide as much as three-quarters of the matter returned to the interstellar
medium (ISM) \citep{Gehrz1989}. During their phase of high mass loss, the envelopes
become opaque at optical and near-IR wavelengths; their hot, innermost
parts are then best studied in the mid- and far-IR and their cool outer
parts at millimeter wavelengths.

%figure 1
\begin{center}
\begin{figure*}
\includegraphics[angle=0,scale=0.78]{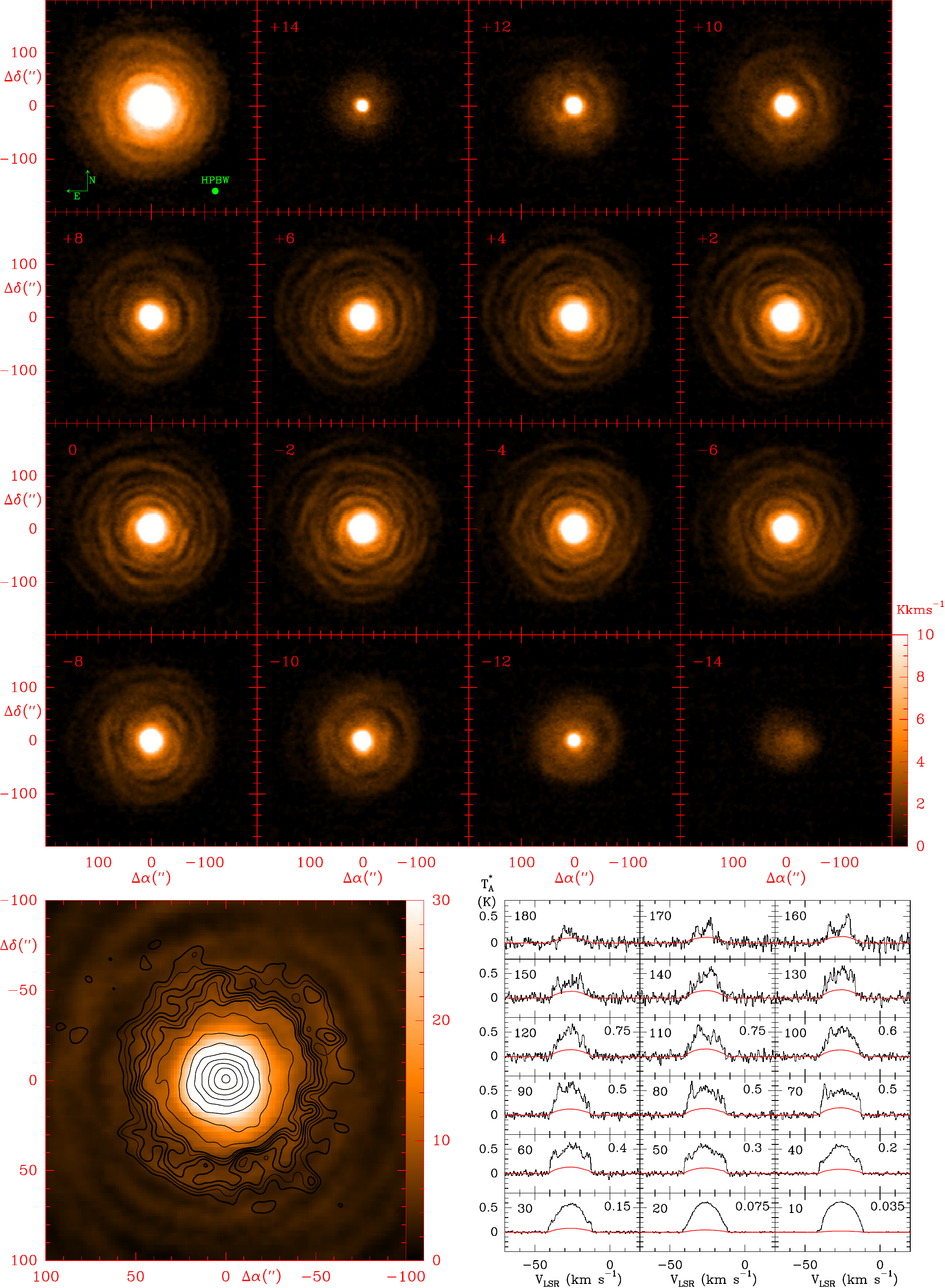}
\caption{{\it Top left box}: Velocity-integrated CO (J=2-1) line emission in the central $400''\times 400''$ 
area ($V_{\rm LSR}$ range from -41 to -12\,km\,s$^{-1}$). {\it Other upper maps}: Velocity-channel maps (resolution 
2 km\,s$^{-1}$); marked velocities are relative to the LSR systemic star velocity ($V_*=-26.5\,$km s$^{-1}$); 
the units of the color scale correspond to K\,km\,s$^{-1}$. Positive and
negative velocities correspond to the rear and front parts of the envelope, respectively.
{\it Bottom right panels}: Observed $^{12}$CO(2-1) line profiles along a strip to the east 
for $\Delta\delta$=0$''$; 
the upper left numbers indicate $\Delta\alpha$ in
arcseconds and the upper right numbers the intensity scaling factor applied to plot all spectra at the same 
$T_A^*$ scale (in K). The contribution of the telescope error
beam has been removed from the data and is shown in red for each spectrum.
{\it Bottom-left panel}: the $^{13}$CO (2-1) line emission integrated between V$_{LSR}$ -28.5 and -24.5 km s$^{-1}$ 
(black contours) superimposed on the $^{12}$CO emission in the same velocity range. Note that the $r=50''$ shell 
also appears in $^{13}$CO. The inner shells are not resolved by the 11$''$ beam. 
}
\label{fig1}
\end{figure*}
\end{center}

IRC+10216, the thick dusty envelope of CW\,Leo and the TP-AGB star the
closest to the Sun, is a choice target for these studies. 
%It was discovered in 1967 by the Two Micron Sky Survey conducted
%by \citet{Neugebauer1969}. 
Its C-rich nature was pointed out by
\citet{Herbig1970} who assigned a cool N type for the star (C9.5).
Besides its
proximity ($\simeq 130$ pc), IRC+10216 is remarkable in several respects: a
large mass ($\sim 2 M_\odot$) and apparent size ($6'$), a nearly
spherical shape and constant expansion velocity, except near the
star, and, last but not least, an incredible wealth of molecular
species: half of the known interstellar species are observed in its
C-rich outer envelope. These molecules
%\LEt {please specify what they refers to here} 
range from CO, 
%\citep{Solomon1971} 
the main tracer of the cool
molecular gas and other diatomic and triatomic species \citep[see e.g.][]{Cernicharo2000,Cernicharo2010}, 
to refractory element molecules \citep{Cernicharo1987}, 
%%metal-cyanides and isocyanides, 
the last detected molecule being HMgNC \citep[]{Cabezas2013},  
to long carbon chains species C$_n$H 
\citep[and references therein]{Cernicharo1996,Guelin1997}, 
%%silicon and sulfur carbon chains SiC$_n$ \citep{Thaddeus1984,Cernicharo1987bis,Cernicharo1989a,Ohishi1989,Apponi1999}, 
%%phosphorus-bearing species \citep[and references therein]{Agundez2008}, 
and include all known interstellar anions (C$_{2n}$H$^-$, n=2,3,4 and C$_{2n+1}$N$^-$, n=0,1,2,  
\citet{%%McCarthy2006,
Thaddeus2008,Cernicharo2008} and references therein). 

In addition to this
large list of molecules expected in a carbon-rich environment, oxygen-bearing species such as H$_2$O \citep{Melnick2001},
%%Hasegawa2006,Decin2010
OH, 
%%\citep{Ford2003}, 
and H$_2$CO \citep{Ford2003,Ford2004} have been also detected. 
While the presence of H$_2$O and H$_2$CO in more evolved C-rich protoplanetary nebula 
%%was already reported \citep{Cernicharo1989b,Herpin2000}, and 
can be easily explained through photochemical processes \citep{Cernicharo2004},
their presence in IRC+10216, a star still in the AGB phase, is particularly difficult to 
explain. 
%%\citep{Agundez2006}. 
\citet{Agundez2010} 
have shown that
an important parameter for the chemical models is the spatial structure of the gas in the circumstellar envelope (CSE) 
which, if clumpy enough, could permit the interstellar UV radiation to penetrate into the inner zones of the CSE and 
trigger a chemistry out of thermodynamical equilibrium.

Although at low angular resolution the mm-wave dust and CO emissions of IRC+10216
appear fairly smooth and spherically symmetric, high resolution
V-light images of the envelope \citep{Crabtree1987,Mauron1999,Mauron2000,Leao2006}, and
emission maps of reactive molecular species
\citep{Guelin1999,Trung2008} reveal multiple shell-like structures
that seem to reflect recurrent episodes of high mass loss with timescales of hundreds of years. The
shells appear patchy and broken into pieces, but are fairly spherical
in the $r=15-30''$ radius region.
%%\citep{Bieging1993,Gensheimer1995,Guelin1993,Lucas1995}.
Surprisingly, they are not exactly centred on CW Leo, but a couple of arcsec away
\citep{Guelin1993}. Emission from reactive species is weaker at
$PA\simeq +20^\circ$ and $-160^\circ$ \citep{Lucas1995}, suggesting
that CW Leo may be leaving the AGB and is developing a bipolar outflow
that sweeps away parts of the slowly expending spherical
envelope. Further evidence of matter ejection along the
$PA=-20\,-\,+160^\circ$ axis comes from the SiS brightness contours
\citep{Lucas1995}, which show that the $4''$ diameter envelope core, as traced by
SiS molecules is elongated in that direction.

At small angular scales the images obtained by \citet{Kastner1994} and
\citet{Skinner1998} show deviations from symmetry suggesting the presence
of an overall bipolar structure. Moreover, high angular resolution
observations of the continuum indicate the presence of clumps and show that the innermost
structures at subarsecond scale are changing on timescales of years
\citep{Haniff1998,Monnier2000,Menshchikov2001,Tuthill2000}. All these fast variations
in  recent 
%%mass loss history of CW Leo 
years led several authors, following \citet{Lucas1995}, to suggest that
%%the star 
CW Leo is beginning its evolution towards the protoplanetary nebula phase
\citep[and references therein]{Tuthill2000}.

More recently, \citet{Decin2011} reported Herschel/PACS maps of
the dust FIR emission showing that the dusty shells observed in
V-light extend in the envelope up to radii $R \geq 3'$. Although 
it was known, from low-resolution studies, that CO emission extends at least
as far, little was known so far about the gas 
distribution at such radii,
more particularly in the region where CO is photodissociated by interstellar (IS) UV
radiation. Moreover, 
contrary to dust, CO yields the velocity information needed to reconstruct the 3-D 
envelope structure.
\citet{Fong2003}, who 
%%have 
mapped the emission up to 
$R=150"$  
%%have 
with BIMA\ (13" synthetized beam), reported the presence of a number of 
arc-shaped features apparently associated with the thin dust shells 
%%as already pointed out in the comparison of molecular emission and reflected light 
reported by \citet{Mauron2000}. However, the limited size and low
sensitivity of the BIMA map prevented them from drawing an image of
the full envelope. To derive 
the molecular gas distribution up to at least 4$'$ from the star
and investigate CW~Leo's mass loss history over the last 10$^4$ years,
we have observed with  high sensitivity the 
$^{12}$CO and $^{13}$CO $J= 2-1$ line emissions throughout the envelope,
using the HERA focal plane array receiver 
on the IRAM 30-m telescope. The HPBW of the telescope, $11''$, or 1400 AU, is small enough
to resolve clumps of gas expelled with a 500 yr delay. \footnote{We
adopt a distance to CW~Leo of 130 pc \citep{Agundez2012} and assume a
uniform expansion velocity of 14.5 km\,s$^{-1}$ \citep{Cernicharo2000}.}

%figure 2
\begin{figure}
\includegraphics[angle=-90,scale=.4]{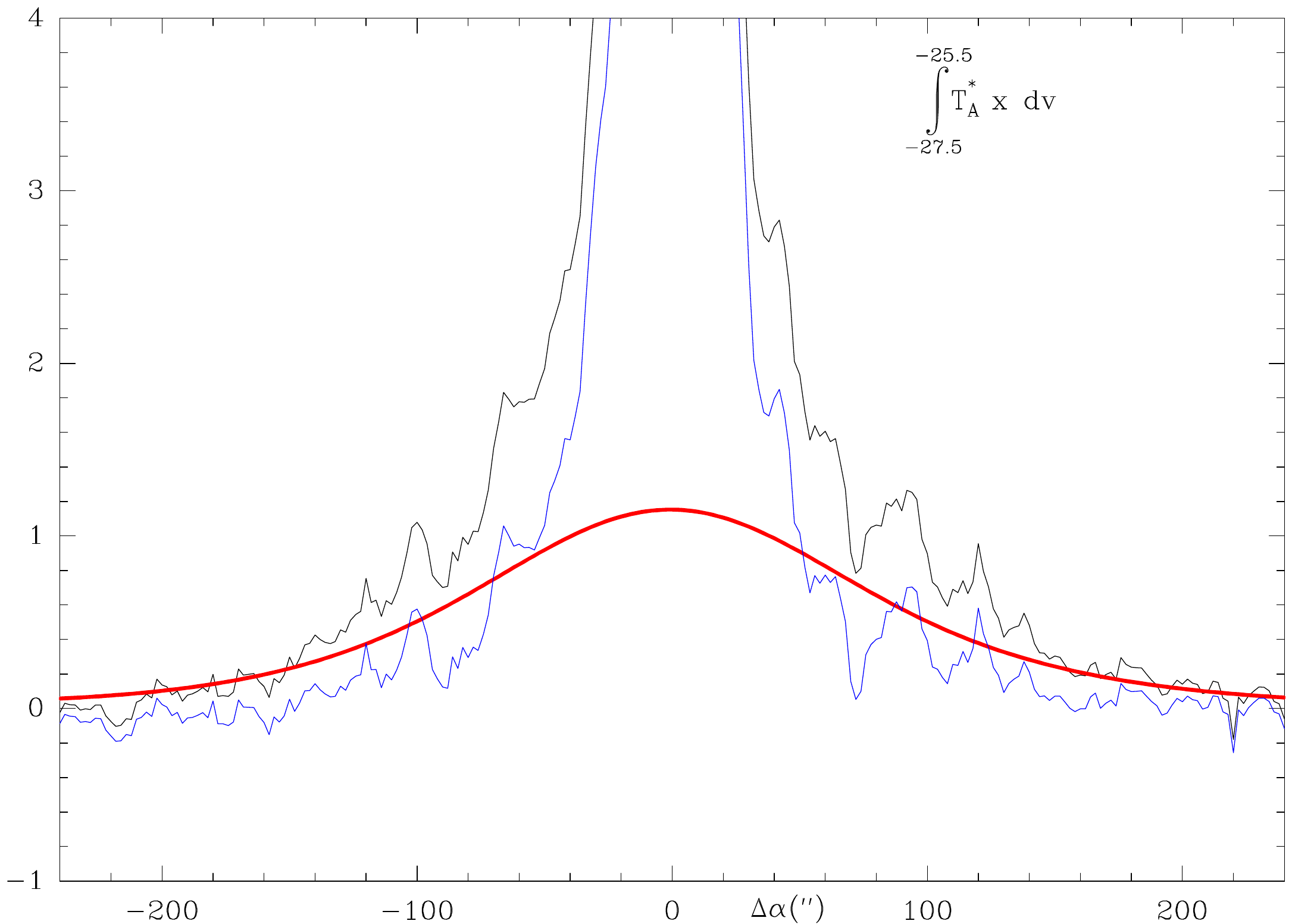} %scale 0.85
\caption{{\it Black line:} Intensity of the $^{12}$CO(2-1) line integrated
between LSR velocities -25.5 and -27.5 km\,s$^{-1}$ observed along an EW strip at the declination of the star
($\Delta\delta=0''$). {\it Red line:}  the response of the telescope error beam to the 
$^{12}$CO(2-1) emission along the same strip. The error beam consists of 3 Gaussians 
of FWHP 65$''$, 250$'',$ and 860$''$ and intensities 1.9\,10$^{-3}$, 3.5\,10$^{-4}$, and 2.2\,10$^{-5}$ 
relative to the main beam, respectively. {\it Blue line:} The $^{12}$CO(2-1) line intensity after removal of the error beam response.
%%(a color version of this figure is available in the online version of the Journal)
}
\label{fig2}
\end{figure}

\section{Observations}
The main observations of $^{12}$CO and $^{13}$CO J=2-1 (see Figure~\ref{fig1}) 
were carried out in February 2009 with the HERA 
9-pixel, dual-polarization receiver on the IRAM 30-m telescope.
Further observations of $^{12}$CO and $^{13}$CO J=1-0 and 2-1 were made between 2011 and 2012, 
using the Eight MIxer Receiver(EMIR): E90 and E230 single-pixel dual polarization receivers.
%EMIR \LEt {Please ensure all abbreviations are written out on first mention throughout, as in:\ 
%Eight MIxer Receiver(EMIR):\ }

The astronomical signals are expressed in the
$T_A^*$, scale (antenna temperature corrected for spillover losses
and atmospheric opacity) in Figures 1 and 2, and in the $T_{MB}$ scale in the subsequent figures
\citep[$T_{MB}=T_A/B_{eff}$, where $B_{eff}$, the telescope beam efficiency is 0.79, 0.78, 0.61, and 0.59
at 110 GHz, 115 GHz, 220 GHz, and 230 GHz, respectively:][]{Kramer2013}.
%\citet[][]{Kramer2013}).
%\LEt {Please remove the parentheses around  "2013" and in any other case when the year is within parentheses.}
The atmospheric opacity was measured from the comparison of
the sky emissivity, recorded every five minutes, to that of hot and cold
loads using the ATM code \citep{Cernicharo1985,Pardo2001}.The zenith sky opacity was 
typically 0.1-0.2.
The spectral data were processed by an autocorrelator consisting of
461 channels spaced by 312 KHz (0.41 km\,s$^{-1}$ at 230.5 GHz).

The nine sky-channels of HERA of each linear polarization were
successively tuned to the frequencies of J=2-1 $^{12}$CO and J=2-1 $^{13}$CO
lines. The receivers operated in single sideband mode, with image
rejections $\geq$ 10 dB.
The observations were made in the
On-The-Fly mode by mapping nine partly overlapping areas of size
$240''\times 240''$. 
Each map consisted in constant declination scans,
separated in declination by 4$''$ (for the central map) or $6''$ (for the other
maps). The scanning velocity was $8''$ per second and the data dumped
every 0.15 seconds, yielding one spectrum every 1.2" in RA. The
observing time per individual map was $\simeq 1$h.

The first map was centred at 0,0, i.e. on the star CW\,Leo, 
maps 2-5 at (+240$''$,0) (-240$''$,0) (0,+240$''$) and (0,-240$''$). Maps 6-9,
of size 120$''$x120$''$, were centred at (+180$''$,+180$''$) (+180$''$,-180$''$) (-180$''$,+180$''$),
and (-180$''$,-180$''$). The combined map fully samples a square area of size
480$''$x480$''$, centred on the star.  Four additional 240$''$x240$''$  areas, centred on the four corners
of the square and extending up to �720$''$  from the star, were also observed.

Prior to observing each individual map, the
pointing and focus of the telescope were checked on the quasar OJ287.
In addition, a small {\it
mini-map}  centred on the star, was observed with HERA in the position switching mode for further checks of
calibration and pointing. The coordinates adopted for position (0,0)
are $\alpha_{2000}$=09$^h$ 47$^m$ 57.3$^s$ , $\delta_{2000}$=13$^o$
16' 43.3$''$. The telescope half-power beamwidth (HPBW) was 11$''$ and 22$''$, respectively, for
the $^{12}$CO J=2-1 and J=1-0 lines, and 11.5$''$ and 23$''$ for the $^{13}$CO
J=2-1 and J=1-0 lines.

The relative positions of the $9\times 2$ sky-channels of HERA were
measured on the first (0,0) map. There, the star and its surroundings
appear in CO emission as a bright compact source. We found shifts of
up to 2" from the nominal channel positions. We measured the relative
calibration of the $9\times 2$ channels of HERA; differences of up to
20\% were found. The shifts and correction factors were applied and
the data re-sampled on a common grid with $\Delta\alpha=\Delta\delta=2''$
before merging the different sky and polarization channels into a single map.
The r.m.s
noise per 0.4 km\,s$^{-1}$-wide channel, after averaging overlapping
maps and both polarizations over a grid with points separated every 2$''$, 
varies between 0.14 K in the central positions and 0.06 K at the map borders
($\pm$240$''$,$\pm$240$''$).

The EMIR observations 
consisted of a partial map of $^{13}$CO $J=1-0$ emission,
and of deep integrations of $^{12}$CO $J=2-1, 1-0$ along selected strips. 
The EMIR $^{13}$CO $J=2-1 and 1-0$ and $^{12}$CO $J=1-0$ data were observed in the raster mode and
consisted of maps of  
$120 \times 120''$ and of strips extending up to $\pm 720''$ from the star (reference position was
15 arcmin west from the star). The typical r.m.s noise per channel of 0.4 km\,s$^{-1}$ is 0.06 K). 
These data are used in section 3.1 to derive the gas physical properties through the envelope.

The 30-m telescope response to point sources is not restricted to a
Gaussian beam with (FWHP of 11$''$), but includes near side-lobes and
an error beam. The latter has been measured at 230 GHz and at 115 GHz and contains
25\% and 14\%, respectively, of the energy collected by the telescope, 
and extends over the entire
IRC+10216 envelope. At 230 GHz, it can be approximated by 3 Gaussians with FWHP of
65$''$, 250$'',$ and 860$''$ \citep{
%%Greve1998,
Kramer2013}. 
The signals observed at position (0,0) are then significantly affected by
the response of the error beam to the outer envelope and those
observed in the outer envelope are affected by the compact central
source. Since our map fully covers the bulk of the envelope CO
emission (whose diameter is $\leq 400''$), we were able to correct the response of
the error beam in every velocity-channel at each observed position. The error-beam corrected
velocity-channel maps are shown in Figure~\ref{fig1} for the central $400''\times
400''$ region. The maps prior to error-beam correction may be found in
the appendix of the electronic version of this paper. Figure~\ref{fig2} shows the $^{12}$CO(2-1)
intensities prior to and after this 
correction along a cut at the declination of the central star. The contribution of the error
beam at the frequencies of the J=1-0 lines of $^{12}$CO and $^{13}$CO is much less important 
as this beam is weaker and broader, and
has not been removed from the data.

\begin{figure}
\includegraphics[angle=0,scale=.44]{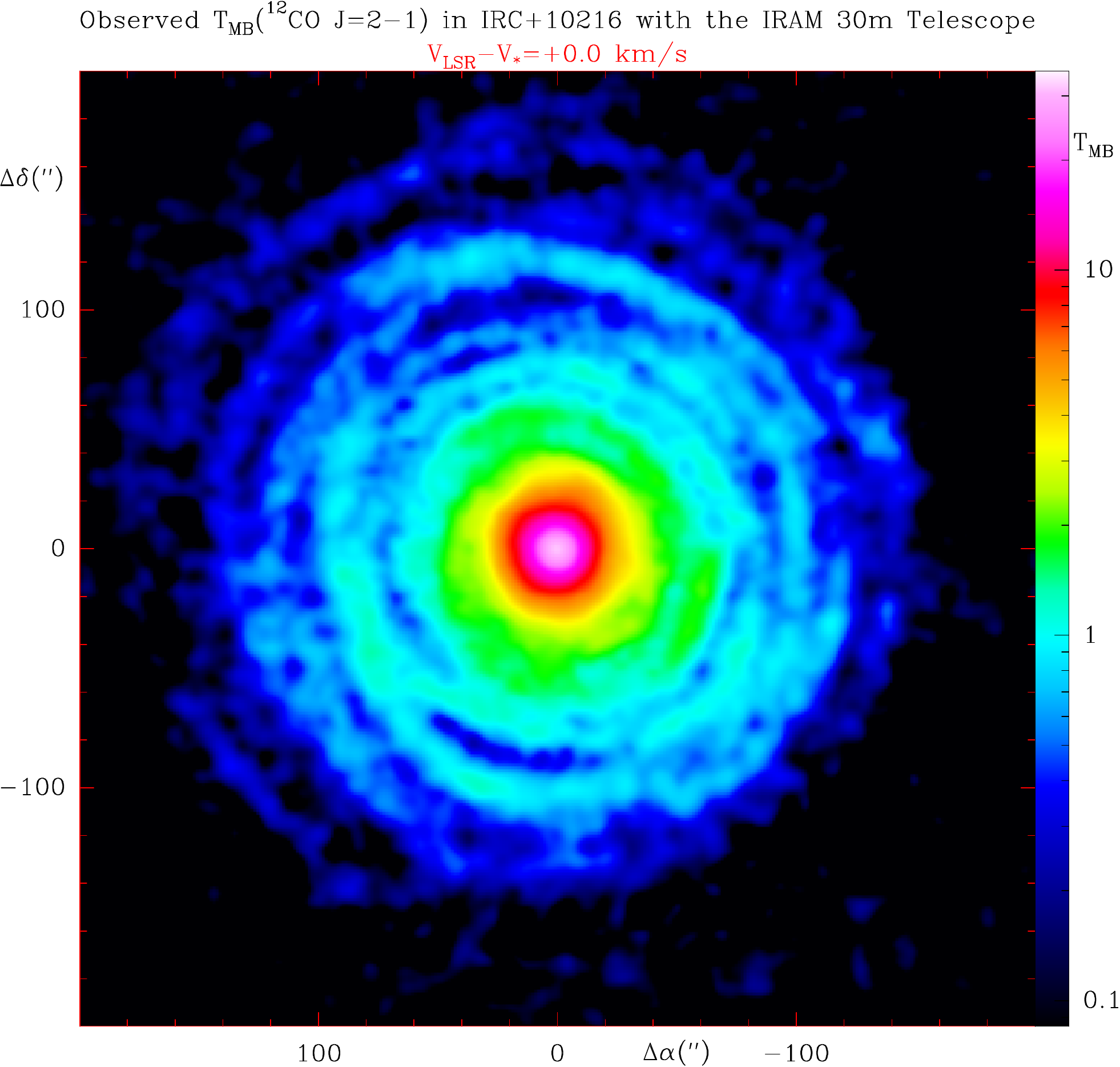} %scale 0.85
\caption{Main beam-averaged $^{12}$CO(2-1) line brightness temperature observed
at V$_{LSR}$-V$_*$=0 km\,s$^{-1}$ ($\Delta$v=0.1 km\,s$^{-1}$).
This is the central frame of an online video showing the distribution of the 
$^{12}$CO line emission at
different velocities.}
\label{fig3}
\end{figure}

Finally, Figure~\ref{fig3} shows
the central frame of a video (online) showing the spatial distribution of the CO emission at different
velocities. The CO data have been re-sampled to a velocity resolution of 0.1\,km\,s$^{-1}$, and hence, there is a 
significant correlation between consecutive frames in the online video.

\section{Results and discussion}
 Given that the outer envelope is in expansion with a constant velocity of
14.5 km\,s$^{-1}$, each velocity-channel map of Figures~\ref{fig1} and \ref{fig3} 
delineates a conical
sector (of opening $\theta$ and thickness $\delta \theta$), whose axis
is aligned with the line of sight to the star CW~Leo. The extreme
velocities correspond to the approaching and receding polar cones and
caps, while the middle velocity ($v=V_{LSR}-V*= 0$ km\,s$^{-1}$, where
$V*$ is the star LSR velocity,$V*= -26.5$ km\,s$^{-1}$) corresponds to a
cut through the star in the plane of the sky.

The multiple shell structure traced by the CO(2-1) emission near $v=0$
km\,s$^{-1}$ is spectacular. It consists of at least seven distinct shells
of radii ranging from $r=45''$ to $r=170''$ and with a fairly high
shell-intershell brightness contrast.  We know from interferometric
maps of reactive molecules \citep{Guelin1999,Trung2008}, as well as 
from V-band images of scattered light,
that more shells or pieces of shell are present at smaller radii: $r\simeq 16''$, $25''$ and $35''$. 
Those are, however, too tight and/or too close to the bright central source to be resolved by the 11$''$
telescope beam. The shells on Figure~\ref{fig1} seem fairly spherical, albeit not
necessarily centred on CW~Leo. Some are locally flattened. 
A nice spiral
structure may be seen at velocities $\pm$8 and $\pm$10 km\,s$^{-1}$ , which suggests,
as discussed below and already indicated by \citet{Guelin1993}, that CW~Leo is 
a binary star with its orbital plane near to a face-on view.
As a
result of off-centring, the shells intersect at several places, possibly a projection effect, tracing a pattern reminiscent of the
rose-window filaments observed by the HST in a number of planetary
nebulae \citep[and references therein]{Balick2002}. 

Obviously, the outer shell pattern teaches us about the mass loss
history during the past $10^4$ yr. It tells us how the envelope
expanded and how far it is penetrated by IS UV
radiation, which is a powerful booster of radical-neutral chemistry.

\subsection{Mass loss history}

We may estimate  the masses of the envelope 
from the line brightness of key molecules in three different ways. 

I) The first method focusses on lines from molecules formed in the upper atmosphere of the star: 
CO, HCCH, HCN \citep{Cernicharo1996b,Fonfria2008}. 
The initial abundances  of those ``parent'' molecules, $x_0$, can be estimated from 
thermochemical equilibrium calculations \citep[see e.g.][]{Agundez2010}. In C-rich envelopes,
the abundance of CO is close to that of oxygen and remains stable up to the CO photodissociation region
(see below). \citet{Agundez2012}
analysed the shapes and intensities of many CO rotational lines, arising from 
widely different energy levels, using parametrized models of the inner envelope and a 
radiative transfer code. They assumed spherical symmetry and used the LVG approximation. 
For $x_0$=CO/H$_2= 6\,10^{-4}$, these authors derived the physical conditions in the inner envelope. 
Their density profile for $r \leq 10''$ yields an average mass loss rate \.{M}$_{t_0}\simeq 2\,10^{-5}\, M_\odot$ 
for the last 500 yr.  

II) The second method relies on the measurement of the CO
photodissociation radius, $r_{phot}$, defined as the radius where the
CO abundance has decreased by a factor of 2: $x$(CO)=$x_0$/2 
\citep[see e.g.][]{Morris1983,Mamon1988,Stanek1995,Doty1998,Schoier2001}.
It assumes that
the surrounding interstellar (IS) UV field is known. Most authors adopt the IS
radiation field of \citet{Jura1974}, which yields in the absence of shielding
a CO photodissociation rate of $G_0$= 2 10$^{10}$ s$^{-1}$ \citep[see Table
1 of][]{Mamon1988}. The method also assumes that the gas density
decreases as $r^{-2}$ in the outermost envelope layers and is not too clumpy. 
Figure~\ref{fig1} shows that the $^{12}$CO(2-1) emission extends up to $r\simeq 180''$ (or $3.6\,10^{17}$ cm) 
and becomes vanishingly small afterwards. Adopting this radius for the 
photodissociation radius $r_{phot}$, we find a mass loss rate \.{M}$_{t}\simeq 2\,10^{-5}\, M_\odot$ 
\citep[see Table~3 of][]{Mamon1988}.
Obviously, this method yields the average mass loss rate at 
times $t> t_0-r_{phot}/v_{exp}$, i.e. more than $8\, 10^3$ yr ago.

III) The last method is based on $^{13}$CO. It assumes that, for $r<<
r_{phot}$, the $^{12}$CO and $^{13}$CO fractional abundances remain 
constant and close to their
nominal values near the star, $x_0$. This is expected since
the CO molecule is chemically stable because $^{12}$CO efficiently
self-shields against photodissociation and partially shields $^{13}$CO, and because both species
are partially shielded by H$_2$. As a matter of fact, the photodissociation
radii of both isotopologues are similar \citep{Mamon1988}. As for $^{13}$C fractionation, it should 
be inefficient in IRC+10216
since, contrary to PNs (e.g. CRL618, Pardo \& Cernicharo 2007) and detached envelopes (e.g. RScu, \citet {Vlemmings2013}), 
the degree of ionization is known to be low in this envelope 
(see \citet{Mamon1988}, for a discussion of CO shielding, isotopic fractionation, and
selective photodissociation in dense CSEs).
Finally, the chances that thermal pulses have changed the  $^{13}$C/$^{12}$C
ratio during the last 8000 years are low \citep{Nollett2003}. Hence, the
$^{13}$CO/$^{12}$CO abundance ratio is expected to remain constant within $r_{phot}$ and close to the
elemental $^{13}$C/$^{12}$C ratio. Molecular line observations do point towards a 
single value $^{13}$CO/$^{12}C$O=$^{13}$C/$^{12}$C=45$\pm 3$ with no change of the $^{13}$C/$^{12}$C 
 ratio between the inner and the outer parts of the IRC+10216 envelope 
\citep{Kahane1988,Cernicharo1991,Cernicharo2000}. From the $^{12}$CO abundance 
derived through method (I) above, $x_0(^{12}$CO)=$6~10^{-4} $, we derive a $^{13}$CO fractional abundance 
$x(^{12}$CO)=$1.3~10^{-5}$. 

The physical conditions and molecular column densities can then be derived in the outer envelope
by comparing the $^{12}$CO, $^{13}$CO (2-1), and (1-0) line intensities. Within the range of
mass loss rates expected, $^{12}$CO should be optically thick and the $^{13}$CO lines
optically thin (see below). We note that since we are not resolving the smallest structures with our $11''$ beam,
we choose to evaluate the $^{12}$CO and H$_2$ column densities from the optically thin $^{13}$CO line intensities and the
$^{13}$CO/$^{12}$CO ratio quoted above, rather than derive the latter ratio from the 
optically thick $^{12}$CO line intensities and the fitted gas volume densities.

\begin{figure}
\includegraphics[angle=0,scale=0.69]{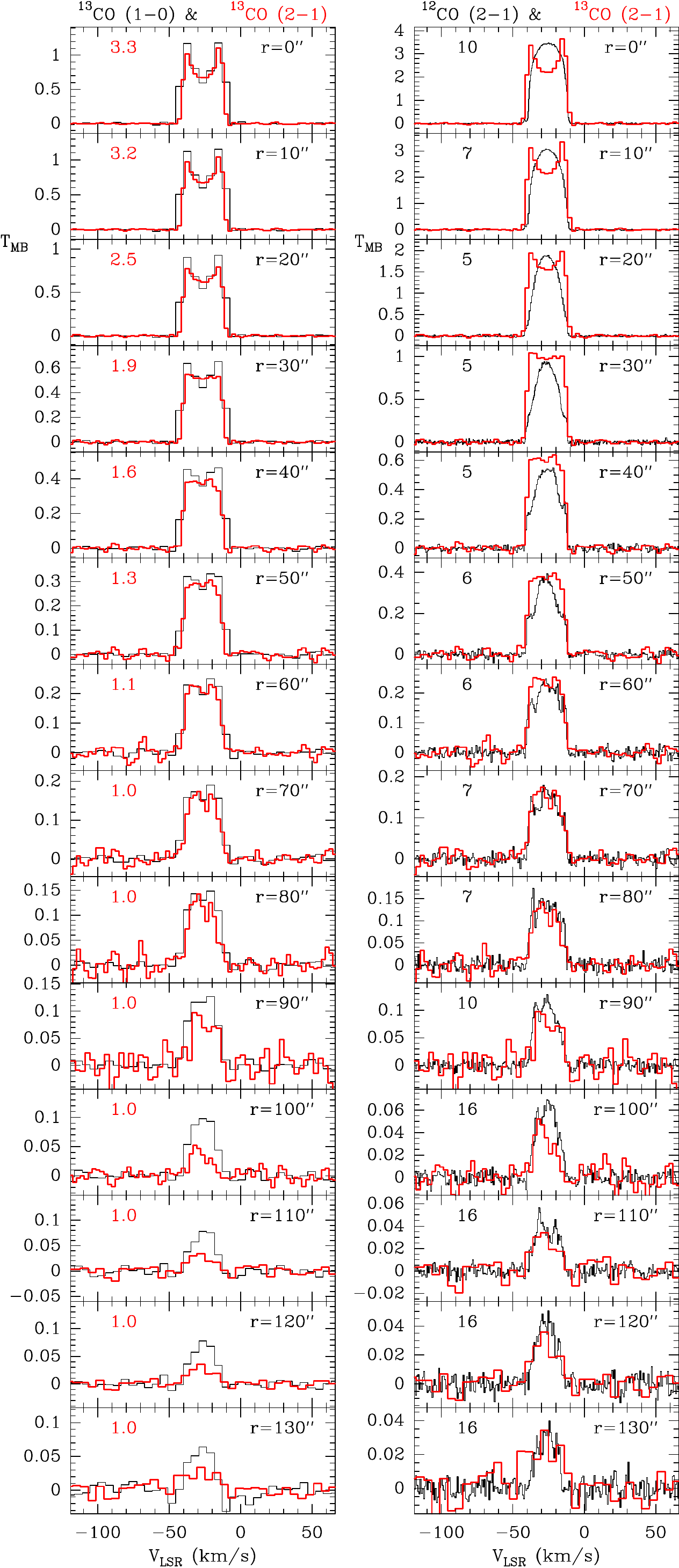} %scale 0.85
\caption{{\it Left panel:} the $^{13}$CO J=1-0 and J=2-1 line profiles  (black and red histograms, 
respectively),
averaged over concentric rings of width  $\Delta$$r =10''$, observed at radii ranging from $r=0''$ to $130''$.
The ordinate scale (black numbers)
applies to the J=1-0 line profile (black). The number in red in the upper left corner
is the factor by which the J=2-1 profile (red) has been divided to match the intensity scale. 
For $r<90''$ it also represents the
$R_{21}=^{13}$CO(2-1)/$^{13}$CO(1-0) velocity-integrated line intensity ratio. 
{\it Right panel:} the J=2-1 $^{12}$CO (black)
and $^{13}$CO line profiles averaged over the same 10$''$-wide rings. The ordinate scale 
%(red numbers)
applies to the $^{13}$CO(2-1) line profile (red). The number in black in the upper left corner
is the factor by which the $^{12}$CO(2-1) profile (black) has been divided to match 
the intensity scale. For $r \geq 40''$, it also 
represents the $R_i=^{12}$CO(2-1)/$^{13}$CO(2-1) velocity-integrated line intensity ratio.}
\label{fig4}
\end{figure}

Figure~\ref{fig4} shows the line profiles observed between $r=0$ and $r=130''$ along a strip at the declination of the star. 
To increase the signal-to-noise ratio, the $^{13}$CO(2-1) and $^{12}$CO(2-1) line 
intensities of Figure~\ref{fig1} have been averaged in azimuth over $10''$ wide concentric rings. 
The numbers in the upper
left corner of each spectrum indicate the scaling ratio between the black and red line profiles (e.g. the $^{12}$CO(2-1) and
$^{13}$CO(2-1) profiles). They  give the $R_{21}=^{13}$CO (2-1)/$^{13}$CO(1-0) and $R_i=^{12}$CO(2-1)/$^{13}$CO(2-1)
velocity-integrated intensity ratios directly, where the black and red line profiles have similar shapes and intensities 
(i.e. at radii $r \leq 90''$ for the left panels and $r\geq 50''$ for the right panels).

We have fitted with a $LVG$ radiative transfer code included in the MADEX code
\citep{Cernicharo2012}, the intensities of the $^{12}$CO and $^{13}$CO J=2-1 and 1-0 
%\LEt {Do you mean CO, CO\ (2-1), and (1-0)?\ Check above as well.   \0}
lines at different radii. We find that the kinetic gas temperatures $Tk$ derived 
for the successive rings of Figure~\ref{fig4} are compatible with the temperature profile $T_{gas}\propto r^{-0.54}$ 
given by \citet{Agundez2006}, albeit $\simeq 20$\% higher. 
>From our multi-line analysis, we find for example at $r= 40'', 60''$, and $80''$ (corresponding
to evolution times of 1700, 2600 and 3400 yr, respectively), $n(H_2)\simeq$ 5\, 10$^3$, 2.2\, 10$^3$ 
and 1.2\ 10$^3$ cm$^{-3}$, $Tk \simeq 20$ K, 16 K, and 13 K.  The resulting mass loss rates are all
\, {M}$_{t}\simeq$ 4\, 10$^{-5}$ $M_\odot$, with an 
uncertainty of a factor of $\simeq 2$ (mainly resulting from the temperature-density near 
degeneracy and the uncertainty on the error beam contribution at large radii, from the difference in 
beamsizes at 115 GHz and 230 GHz at small radii, and from the uncertainty on the $^{13}$CO 
fractional abundance). For $r=120''$ we find ${M}_{t}= 3.5\, 10^{-5}$ $M_\odot$. 
These mass loss rates are within the uncertainties equal to those quoted above 
for $r< 10''$ and $r> 180''$.

Our study shows that the $^{12}$CO(2-1) line is very thick at small radii, and moderately thick ($\tau=4-2$) at radii larger
than 40$''$. The $^{13}$CO lines are optically thin ($\leq 0.1$), and the (2-1) $^{12}$CO and $^{13}$CO lines sub-thermally 
excited at radii $r\geq 40''$. The (2-1) line excitation becomes very low for $r\geq 90''$, where the average $^{12}$CO(2-1) 
line brightness temperature drops below 1K and the $R_{21}$=$^{13}$CO(2-1)/$^{13}$CO(1-0) ratio drops below 1. We note 
that the LVG models fail to reproduce the $^{12}$CO(2-1) line intensities and the $R_i$=$^{12}$CO(2-1)/$^{13}$CO(2-1) 
intensity ratio of Figure \ref{fig4}, both of which are a factor of 2-3 times smaller than predicted for $40''\leq r \leq 80''$. 
This results from an underestimation of the $^{12}$CO line opacity because of the presence of dense sub-structures in the shells.
These structures appear on high-resolution interferometric CO maps 
(Gu\'elin et al. {\it in preparation}), as well as
on the visible-light images \citep[e.g.]{Leao2006}. 
They are unresolved by the $11''$ telescope beam and smoothed out by our azimuthal averages. 
They probably contain a fair fraction of the total gas mass and significantly contribute to the optically 
thin $^{13}$CO emission, albeit they contribute little to  the optically thick $^{12}$CO(2-1) emission, which is damped. 
The relatively low $^{12}$CO intensities could also result from a decrease of the $^{12}$CO/$^{13}$CO abundance 
ratio at radii $\geq 40''$; however, selective photodissociation and thermal pulses, which are the mechanisms the most 
likely to affect this ratio, should increase, rather than decrease the $^{12}$CO/$^{13}$CO ratio, making this 
explanation unlikely. Finally, although the J=2-1 CO lines are much less sensitive to radiative IR pumping  
than the J=3-2 and 1-0 lines, the effect is not negligible and may affect both $R_{21}$ and $R_i$ 
\citep[see e.g.][]{Cernicharo2014}. 
Even though the observed values of the $^{12}$CO/$^{13}$CO intensity ratio are low, they follow the 
predicted behaviour: high at small radii, where $^{12}$CO(2-1) is optically very thick, they are a minimum of between 
$r= 30''$ and 60$''$ and strongly increase at $r>100''$. As for the 
$R_{21}$ line intensity ratio, it is accurately reproduced by our LVG calculations.

%figure 4

The result of our investigation is that the mass loss rate, averaged
over periods longer than the periodic modulation that gives rise to the dense
shells, seems within a factor of 2 constant over the last 10$^4$
years. Integrated over the range $r_*\leq r \leq 180''$, it yields an
envelope mass $M(r_{phot})
%%=0.16
=0.3\, M_\odot$. The total mass lost by the
star since it joined the AGB phase can be estimated to be $\simeq 2\, M_\odot$
%\LEt {Do you mean?\ can be estimated to be much larger than the first shell...?} 
from the first shell ejected 70000 yr ago \citep{Sahai2010}.
%%$\simeq 1.4\, M_\odot$.

Prior to our investigation, several attempts have been made to estimate the mass
loss rate and the envelope mass. 
\citet{Crosas1997} and \citet{DeBeck2012}
used mm and sub-mm lines of $^{12}$CO and $^{13}$CO, observed 
towards CW~Leo or near this star, to derive 
a mass-loss rate averaged along the line of sight. They found values within a factor of 2
equal to ours. Other attempts were based on the dust thermal emission. 
\citet{Decin2011} derived from the Herschel/PACS 100$\mu$m
flux, after removing the background emission, an envelope mass of 0.23
$M_\odot$ inside a 390$''$ radius. For a distance of 130 pc and an
expansion velocity of 14.5 km\,s$^{-1}$, this corresponds to an average
mass loss rate \.{M}=1.4 $10^{-5}$ M$_\odot$\,yr$^{-1}$ over the last 1.7 10$^4$ yr. Those
values, however, depend heavily on the assumed dust temperature
profile, since the 100 $\mu$m dust emission varies as $T_d^{3-4}$ (100
$\mu$m is close to the peak of the dust emission for $T_d\simeq
30$K, the average temperature in the outer envelope). 
They also depend on the dust emissivity coefficient and
dust-to-gas mass ratio (assumed equal to 250), two parameters 
uncertain by factors of a few. Obviously, they depend on the 
accuracy of the 100 $\mu$m flux measurement, which are uncertain by
a factor $\geq 2$ because of PSF artefacts \citep{Decin2011}. 
In view of all those uncertainties, the mass loss rate derived
by Decin and coworkers is remarkably close to that we derive from
CO.

Neither the high quality and high resolution CO data reported here and the
FIR data from Herschel/PACS support earlier findings, based on
balloon or IRAS FIR data, that the average mass-loss rate has strongly
decreased in the past few thousand years (by a factor of 2 in the past 2000 yr,
according to \citet{Fazio1980}, and by a factor of 9 in the past 7000 yr,
according to \citet{Groenewegen1997}.
>From an analysis of
all the CO data available in 1998, using an elaborated source and
radiative transfer models, \citet{Groenewegen1998} also advocated a
decrease of \.{M} by a factor of 5 in the last 3000
yr. However, as pointed out by these authors, the IRAS and balloon
observations lacked either angular resolution or 
%\LEt {and/or is considered vague. Do you mean?\ lacked either angular or ...}
spectral coverage,
while the CO data available at that time for the outer envelope were widely discrepant. Our
30-m telescope data and the HIFI and PACS data benefit from a higher
sensitivity and angular resolution. Our CO maps were observed by good
weather, mostly through fast on-the-fly scanning, and are certainly much more
homogeneous and more reliable than those used by Groenewegen and
collaborators. We note that \citet{Teyssier2006}, from a relatively
recent analysis of the $^{12}$CO(1-0) through (6-5) lines, arrived at
\.{M}= 1.2 $10^{-5}$ M$_\odot$\,yr$^{-1}$ (1.8\,10$^{-5}$ M$_\odot$\,yr$^{-1}$ scaled to
our adopted distance of $D=130$ pc, and expansion velocity of 14.5 km\,$^{-1}$), 
a value not much smaller than our value.

\subsection{Interaction with the external medium and UV radiation}

Actually, the IRC+10216 envelope is known to be older than 10$^4$ yr
and to extend further out than shown in Figure~\ref{fig1}. We have used the IRAM
30-m telescope for longer integration observations along a NS
line extending up to $360''$ N of the star. We did detect CO emission
at very low level, but only at $r=240''$; we could not check, however,
whether this emission corresponds to a detached shell.
In the GALEX UV images, Sahai and Chronopoulos (2010)
discovered  a bright semi-circular rim, 12$'$
in radius centred some 4$'$ W from the star. They assign the rim to
the front of the shock at the interface between the stellar wind and the surrounding
gas. The 12$'$ radius already suggests an age $> 3\, 10^4$ yr. From an
analysis of the rim shape, Sahai and Chronopoulos (2010) derive an age
two times larger (6.9 $10^4$ yr).

Inside the FUV rim, the stellar wind should be freely streaming so
that the shapes and motion of the expanding shells observed in the CO and dust 
emissions teach us about the way matter has been expelled from the stellar 
atmosphere. The CO distribution in the outermost shells must also be re-shaped by 
interstellar UV and teaches us about the ambient UV field. 

Figure~\ref{fig1} shows that the envelope is dissymmetric and extends farther in the N-NE direction. This 
is particularly clear on the maps between v= 0 and -6 km\,s$^{-1}$. South-west should be the privileged 
direction for incoming interstellar UV radiation, as it is the direction of the Galactic Plane. 
The CO asymmetry 
may then result from photodissociation by an asymmetric UV radiation field. Alternately, it may come from the 
ejection mechanism, as discussed in the next section.

%quote here the HI detection
%Quote Juvella 2007: Khi UV 30% smaller

%Figure 5
\begin{figure}
\includegraphics[angle=0,scale=.41]{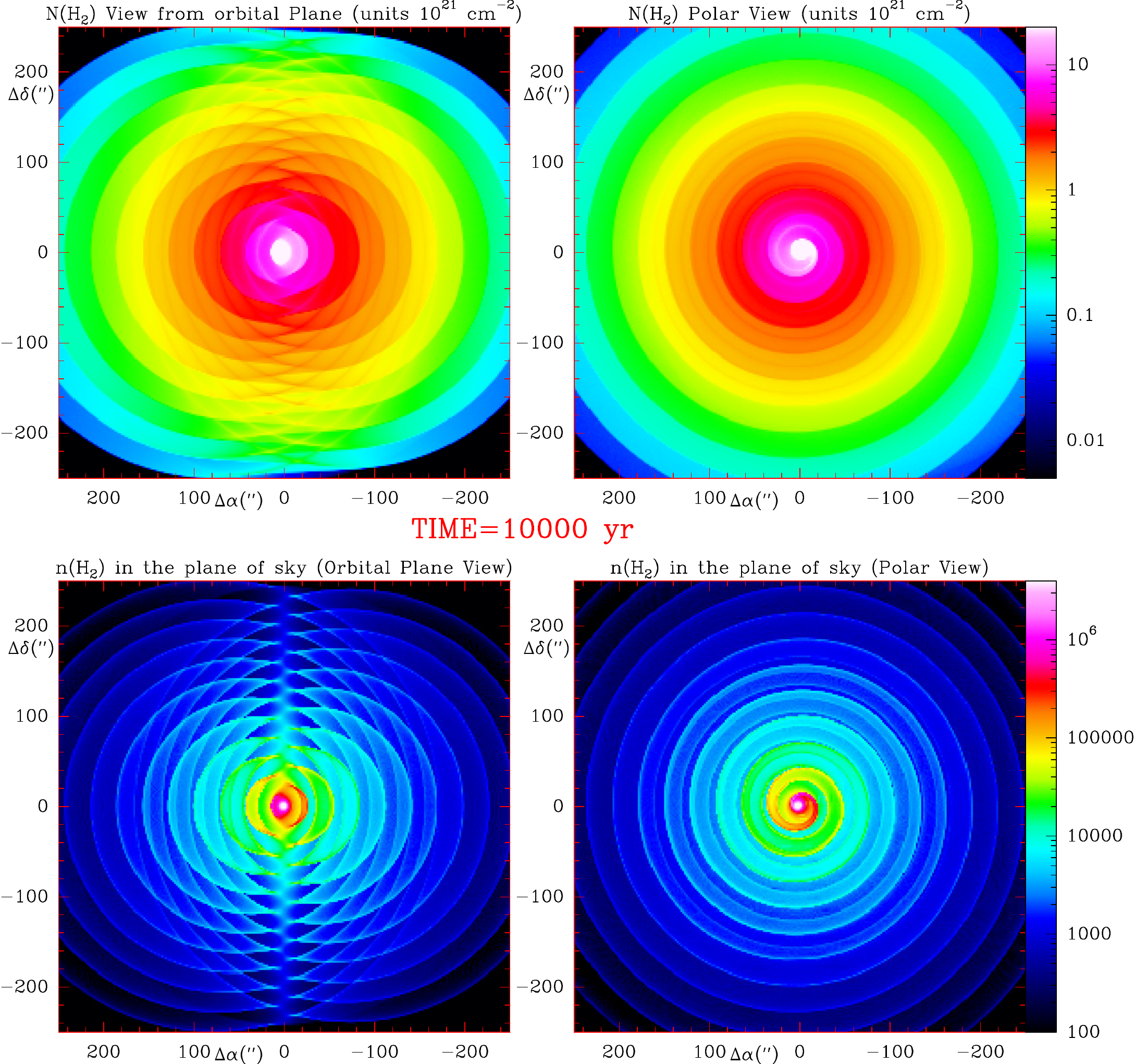} 
\caption{{\it Top Panels:} H$_2$ column density distribution after 10$^4$ yr of constant 
mass loss for the model described in the text. The column density is integrated over all velocities.
The right panel corresponds to a polar view of the CSE (line of sight perpendicular to the 
orbital plane) and the left panel to a view from the
orbital plane. {\it Bottom Panels:} H$_2$ volume density in the plane of the sky for
the same views. The figure corresponds to the last frame of an online
video showing the time growth of the CSE.} 
\label{fig5}
\end{figure} 

\subsection{Mass loss mechanism}

As can be seen in Figure~\ref{fig2}, which shows the $^{12}$CO
velocity-integrated emission profile at the declination of the star,
the shell-intershell brightness contrast is $>3$ for $r>60''$, the
radii where the shells seem resolved by the 30-m telescope beam. \citet{Decin2011}
quote a typical value of 4 for the shell-intershell
contrast of the FIR dust shells. The scattered light images of the
inner region also suggest a similarly large contrast ($3-5$) \citep{Mauron1999,
Mauron2000, Leao2006}. Thus, on the scale of individual shells
(the shell separation is typically $10^3$ yr), the mass loss process is highly
variable. Yet, as we have seen, it seems fairly stable averaged
over longer timescales. Obviously, one must be looking for periodic events.

CW~Leo is known to be a Mira-type variable. Mass loss is supposed to
occur in Mira-type stars when the uppermost surface layers expand and
become cool enough to form dust. Dust grains are accelerated outwards
by the stellar radiation pressure, dragging gas on their way. However,
the pulsation period of CW~Leo, which is 649 days \citep{LeBertre1982}, is much too short to
explain the spacing between the shells. Other periodic surface
expansion events are linked to thermal pulses. Those result from the
periodic ignition of the helium shell that surrounds the stellar
carbon core. However, their period is much too long (few$\times
10^4$ yr, \citet{Forestini1997}). The time separation
between the shells, $0.5-1\,10^3$ yr, implies another 
mechanism.

%Figure 6
\begin{figure}
\includegraphics[angle=0,scale=.41]{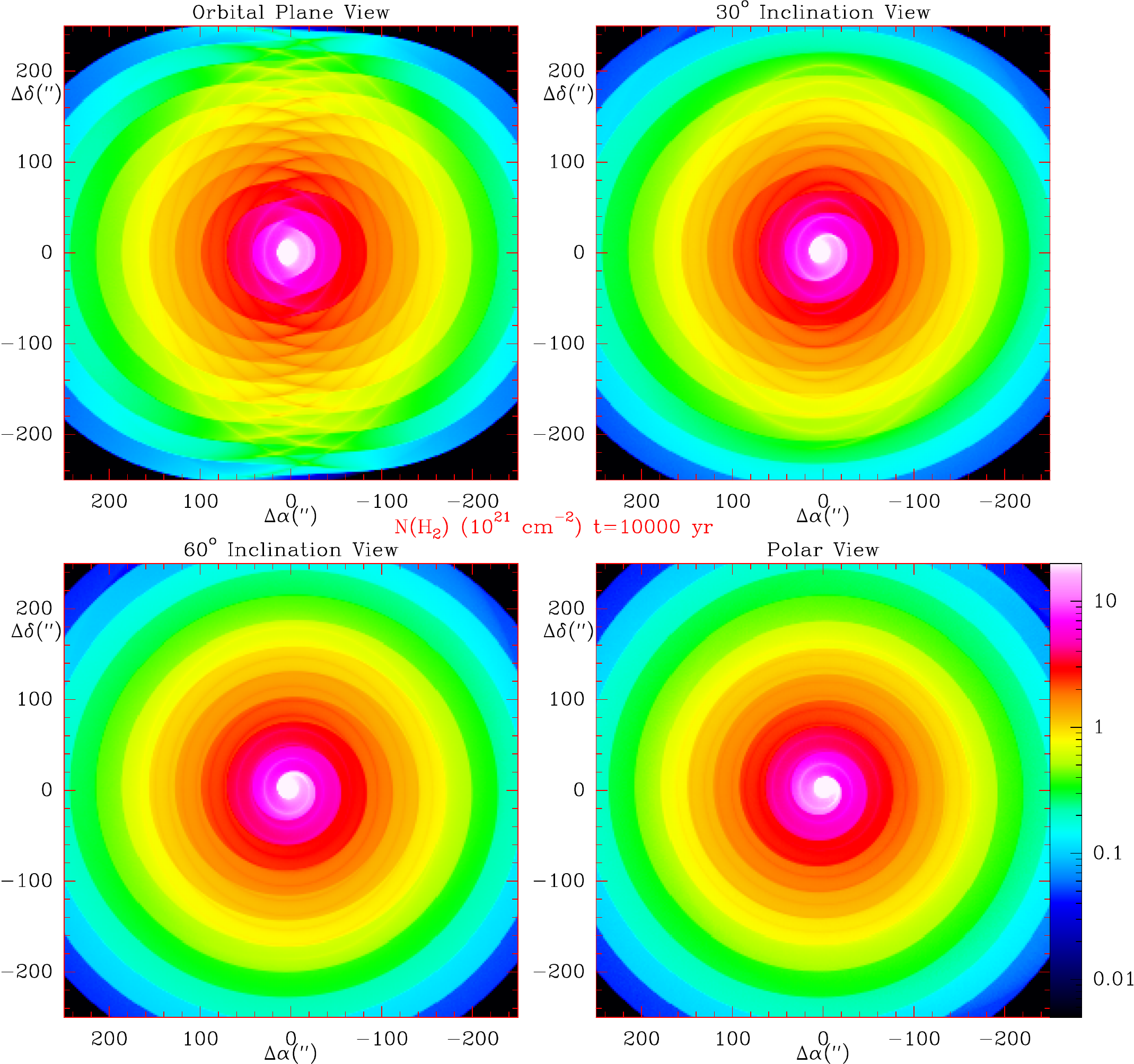} 
\caption{H$_2$ column density distribution, in units of 10$^{21}$ cm$^{-2}$, at t=10000 yr for the same
model as in Figure \ref{fig5}, but for different inclinations of the line of sight with respect to the orbital plane.} 
\label{fig6}
\end{figure} 

%Figure 7
\begin{figure}
\includegraphics[angle=0,scale=.41]{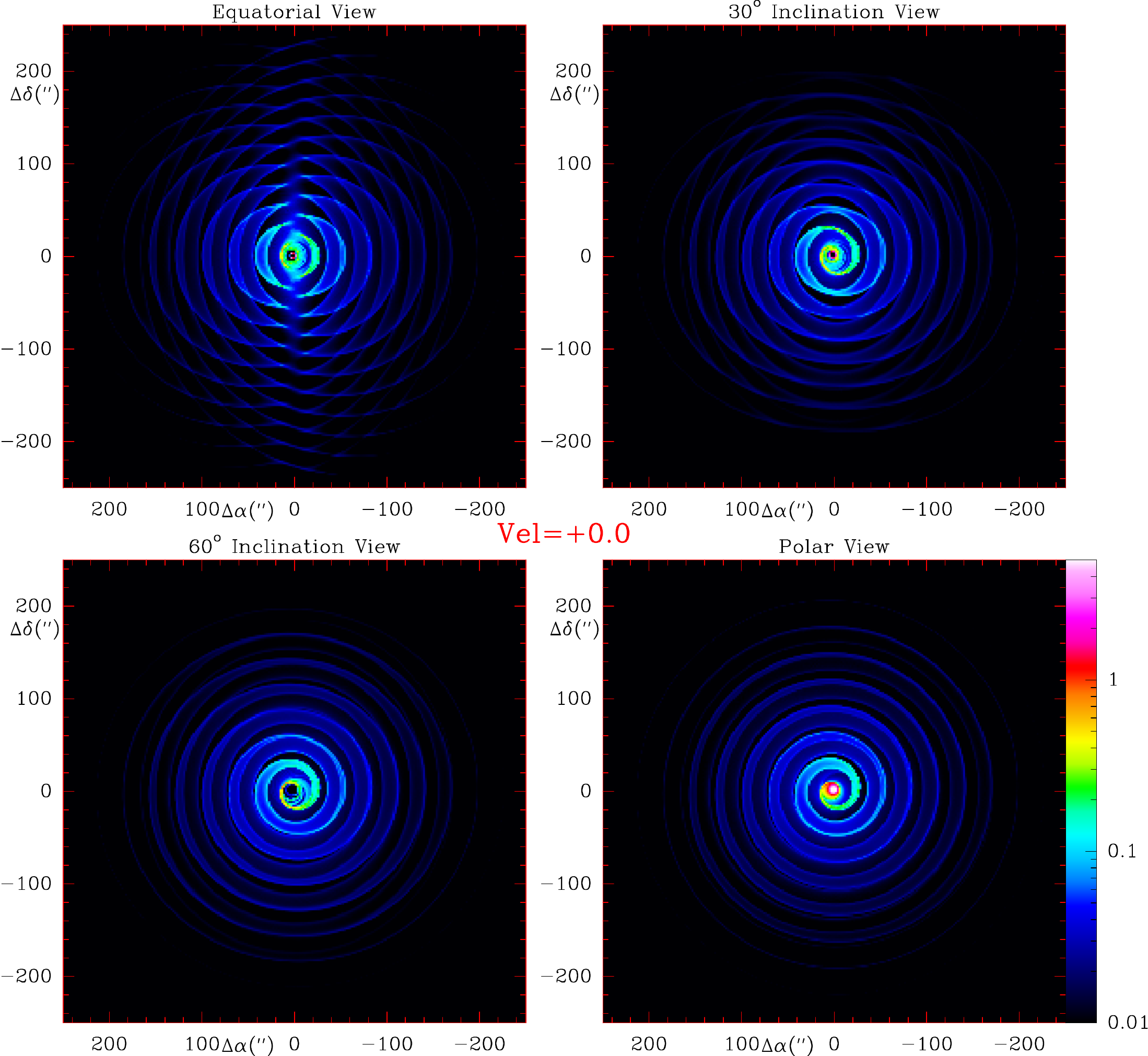} 
\caption{H$_2$ distribution of the gas at the velocity of the star at t=10000 yr. The model parameters are 
the same as in Figure \ref{fig5}. The figure corresponds to the middle frame of an online
video showing the distribution of matter in the envelope at different velocities.} 
\label{fig7}
\end{figure}

The molecular shells of Figure~\ref{fig1} are not exactly centred on CW~Leo. This was
first noticed by \citet{Guelin1993}, who pointed out that the
$r=15''$ shell visible on the $v=0$ km\,s$^{-1}$ CCH and
C$_4$H maps, despite its spherical shape, was centred 2$''$ NW of the star. The authors
suggested the shift results from an orbital motion of CW~Leo. The
motion, caused by the presence of a companion, would impart a 
drift velocity to the material expelled by the star. As a consequence, 
a shell of gas expelled near perihelion will drift 
away from the star in a direction that depends on the orbital phase at
the time of its ejection. The ejection phase may vary from shell to shell
leading to a complex rose-window like pattern similar to
%not unlike 
that of Figure~\ref{fig3}.

\begin{figure}
\includegraphics[angle=0,scale=0.4]{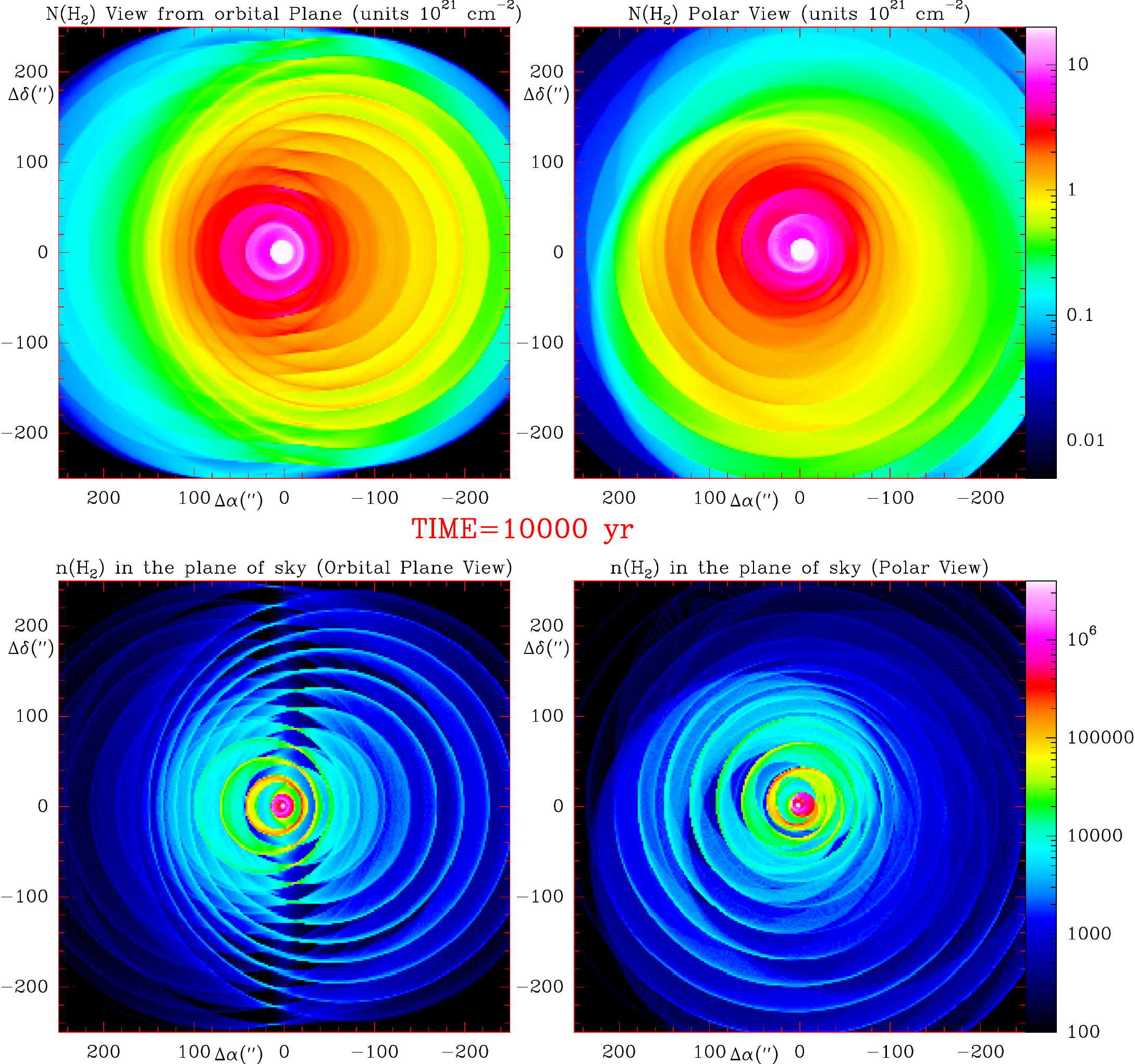}
\caption{{\it Top Panels:} H$_2$ column density distribution after 10$^4$ yr 
derived for the case of a periodic mass loss enhancement caused by the close fly-by of 
a companion star with an orbital period of 800 yr for the model described in the text. 
The column density is integrated over all velocities.
The right panel corresponds to a polar view of the CSE (line of sight perpendicular to the 
orbital plane) and the left panel to a view from the
orbital plane. {\it Bottom Panels:} H$_2$ volume density in the plane of the sky for
the same views. The figure corresponds to the last frame of an online
video showing the time growth of the CSE.} 
\label{fig8}
\end{figure}

\begin{figure}
\includegraphics[angle=0,scale=.41]{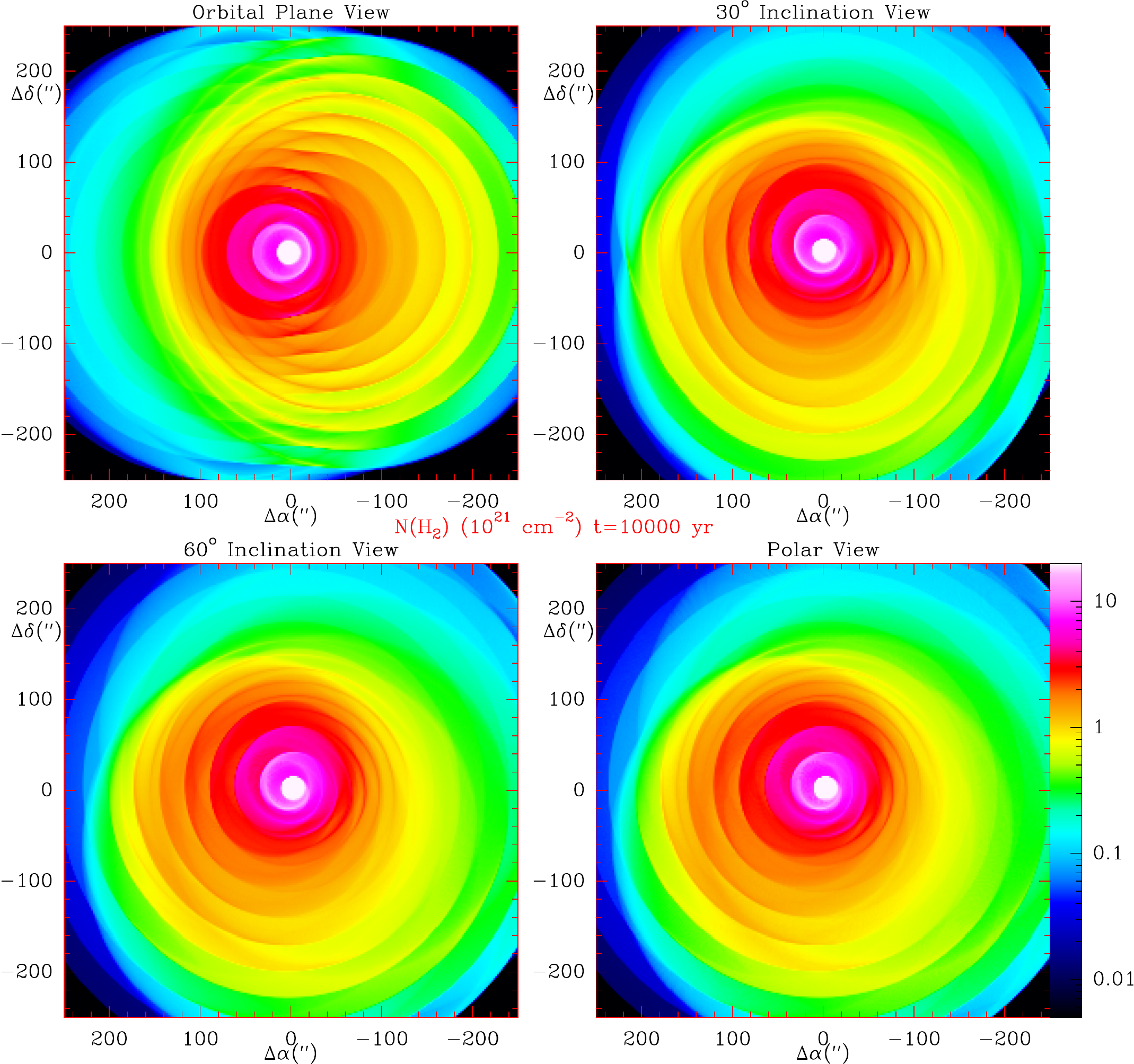} 
\caption{Total H$_2$ column density (in units of 10$^{21}$ cm$^{-2}$) after 10000  yr, for the same
model as Figure 8 at different inclinations of the line of view with respect to the orbital plane.} 
\label{fig9}
\end{figure}

\begin{figure}
\includegraphics[angle=0,scale=.41]{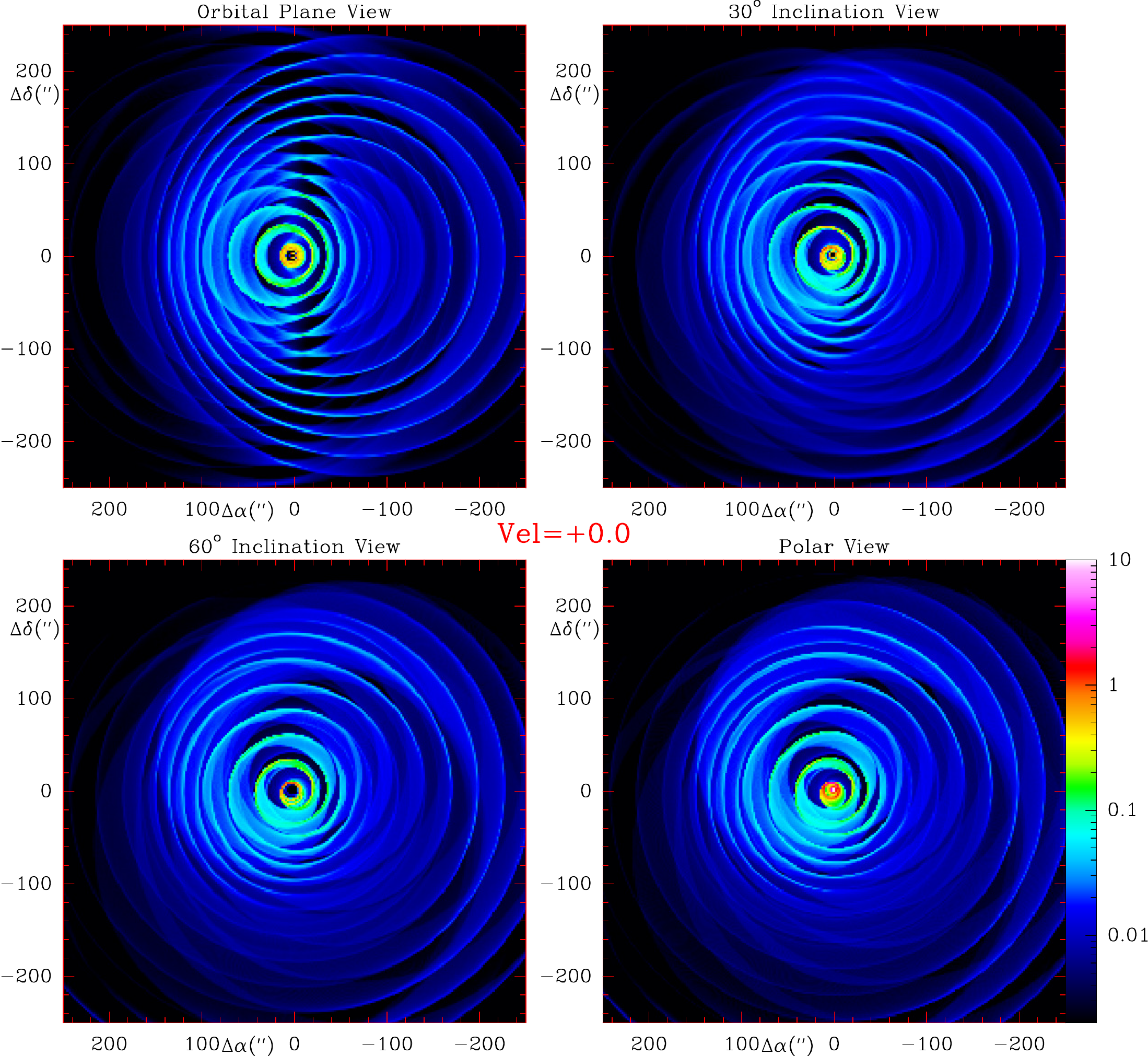} 
\caption{Spatial distribution of matter at t=10000 yr for a velocity of the
gas with respect to the star of 0 km\,$^{-1}$ in the case of periodic mass loss (same parameters as
on Figure 8). The figure corresponds to the middle frame of an online
video showing the distribution of matter in the envelope at different velocities.} 
\label{fig10}
\end{figure}

\begin{figure}
\includegraphics[angle=0,scale=.41]{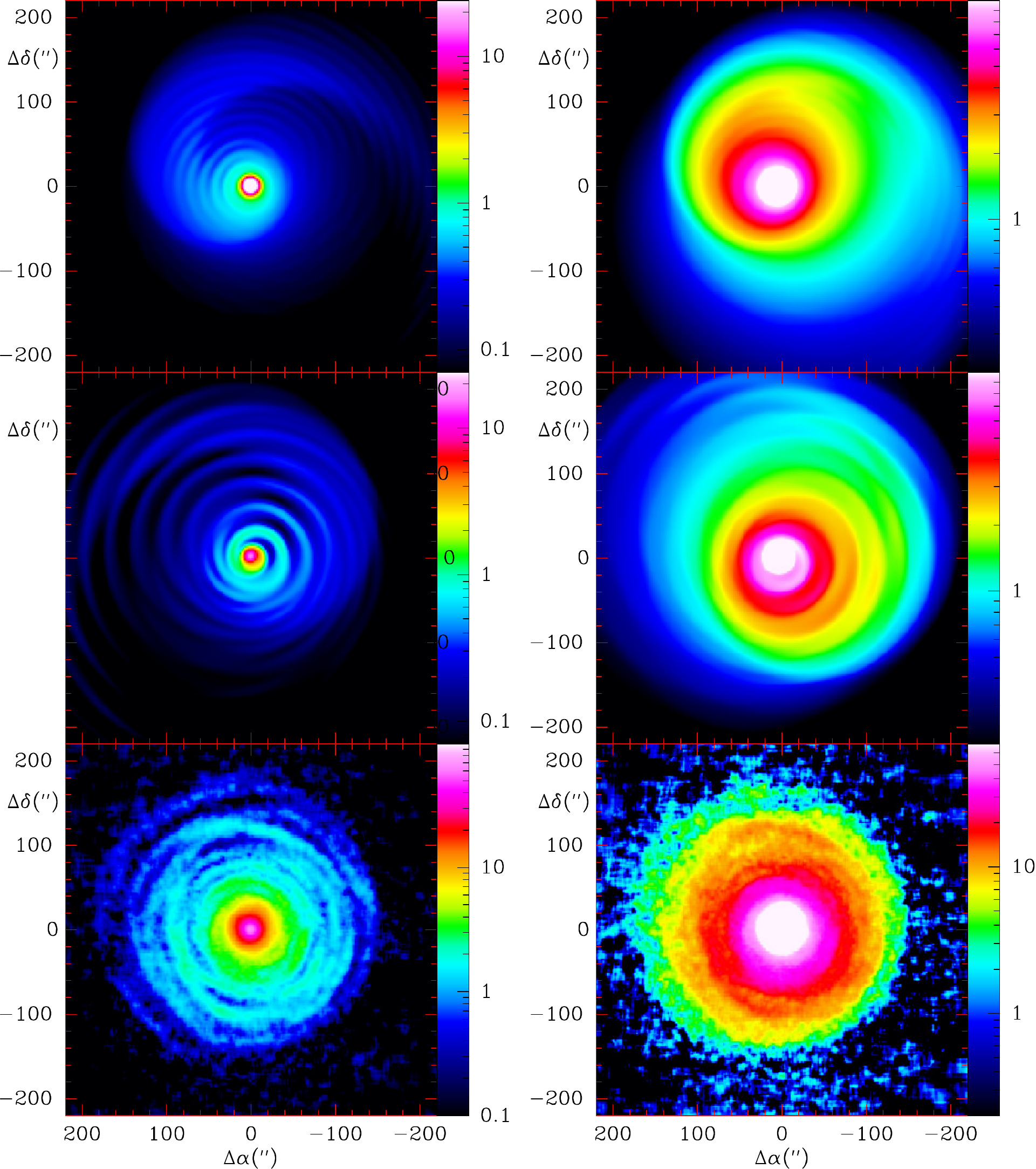} 
\caption{Gas distribution at t=10000 yr for an enhanced mass loss with periods of
400 yr (upper panels) and 800 yr (middle panels), compared to the observed CO(2-1) line 
brightness distribution (lower panel). The model data have been convolved with a Gaussian of width equal 
to the telescope HPBW. The left panels  show the distribution of the 
gas for velocities (V-V$_*$)= $\pm$ 2\,km\,s$^{-1}$ and the right panels the gas column density 
integrated over all velocities.} 
\label{fig11}
\end{figure}

We have attempted to mimic the resulting pattern for a range of 
orbital parameters, using a simple model of the growth of an envelope
around a binary system. 
Our model assumes that the mass loss rate of CW~Leo is enhanced at certain phases 
of the orbital motion of the double star, e.g. near periastron in the case of a
high excentricity orbit. The ejection mechanism, which may be linked to the change of 
the Roche lobe, is not considered by itself and the ejection
of matter is assumed to be spherically symmetrical with respect to CW~Leo. The ejected matter 
is a collection of shells consisting of freely moving particles, with an expansion velocity of 14.5 km/s 
with respect to CW~Leo. 
The time step in our calculations is 2 yr. Consequently, each shell has a thickness of 9.15\,10$^{13}$ cm,
i.e. 1.3 stellar radii.
The gravitational wake of the companion, assumed to be lower in mass than CW~Leo, is not considered. 
The particle motion is followed inside a grid attached to the centre of the 
double star system. A detailed description of the mathematical formalism can be found in 
\citet{He2007}. We have considered a number of cases by varying the total system mass $M_{sys}$, 
the companion-to-CW~Leo mass ratio
(from 0.5 to 1.5), the orbital period ($P_o=400$ to 2000 yr), and the duration of the high mass loss episodes
($\Phi$ from 2$\pi$-0.5 to 0.5 rd). Figures~\ref{fig5} to \ref{fig11} show different views of the 
matter (i.e. molecular gas) distribution in two cases of interest: constant mass loss and mass loss enhanced 
by a factor of 10 
for a fraction of the orbit. In those examples, we assumed $M_{sys}= 3 M_\odot$; to ease the comparison, 
we choose in both cases the same orbit: circular with a radius of 130 AU, and a period $P_0=800$ yr. 
The orbital velocity of CW~Leo is 4.5 km\,s$^{-1}$. 
 
Figure~\ref{fig5} shows the patterns predicted after 10$^4$ yr of constant mass loss rate. The adopted gas density 
at $r_0$ is 3\,10$^{10}$ cm$^{-3}$.
The figure shows a pattern of over-dense ring-like structures shaped by the orbital velocity transmitted by CW~Leo
to each expelled shell of matter. The mass and density distributions are symmetric 
with respect to the orbital plane. The polar view shows a tightly wrapped spiral pattern in the orbital plane. 
This type of spiral structure has been observed in the envelopes of AGB stars AFGL 3068
(Mauron \& Huggins 2006) and, more recently, R Sculptoris \citep{Maercker2012}. 
\citet{Mastrodemos1999},\citet{Kim2012},\citet{Kim2013}, and \citet{He2007} have 
computed several examples of such patterns. We note that the spirals
appear in a natural way in binary star systems without the need of a discontinuous mass loss. The 
$online$ video
associated with Figure \ref{fig5} shows the spatial distribution of mater as
a fonction of time, i.e. during the growth of the envelope, from t=0 to 10000 yr. 
Figure \ref{fig6} allows us to compare the envelope for four different inclinations $i$ of the orbital 
plane with respect to the line of sight. The spiral structure visible for $i=90^\circ$ gets distorted 
when the inclination decreases. The edge-on view of the orbital plane 
shows two series of circles symmetrically shifted with respect to CW~Leo.
To obtain a better insight and constrain the orbit inclination, we have computed the column density of gas, N(H$_2$), 
for velocities from -14.5 kms$^{-1}$ to 14.5 kms$^{-1}$. Figure \ref{fig7} shows N(H$_2$) for $V-V_*$=0 km\,s$^{-1}$ and 
the associated {\it on line} video  N(H$_2$) for all velocities.

%Figure 9

Figure~\ref{fig8} shows the pattern produced by periodic mass loss enhancements. The parameters
are the same as in Figure~\ref{fig5}. The orbital parameters are chosen to be the same as in Figures \ref{fig5} 
through \ref{fig7} to make the comparison with Figure~\ref{fig5} more pertinent. The mass loss 
integrated over one period is the same as in Fig. 5, but the mass loss rate increases by a factor 
of 10 when the orbital phase is within $30^\circ$ from phase 0 (i.e. periastron for an elliptical orbit).

%Figure 10

Figure~\ref{fig8}, i.e. the periodic mass loss simulation, replicates several key features of Figure~\ref{fig1}:
{\it i)}:the presence of a number of over-dense rings in the plane of the sky (see the polar view: lower right figure),
some of which intersect at places, {\it ii)} increasing separation between the bright rings at large 
radii, and {\it iii)} asymmetrical envelope extent in the orbital plane.  
Figure \ref{fig9}  shows the total N(H$_2$) at t=10000 yr for different inclinations of the
orbital plane, and Figure \ref{fig10} and its associated {\it \textup{online}} video, N(H$_2$) as a function of the gas
velocity.
The CO line profiles in IRC+10216, like those of most molecular lines \citep{Cernicharo2000},
show a range of velocities restricted to $V-V_*$=-14.5 to 14.5 km\,s$^{-1}$, with very
sharp edges. This is consistent with a near face-on orientation of the orbital plane: 
A much smaller inclination of the plane to the line of sight would produce a significant
line broadening in the form of line wings because of the change in the star apparent radial velocity 
between the mass loss events.

%Figure 11

Figure 11 shows a comparison between the episodic mass loss model of Fig. 8 and the observations.
The model data have been convolved with a Gaussian of width equal to the telescope HPBW, i.e. 
$11''$. The left panels show the distribution of the gas for velocities (V-V$_*$)=$\pm$\,2\,km\,s$^{-1}$. 
The right panels show the gas column density integrated over all velocities. 
The upper panels correspond to the episodic mass loss model shown on Figure 8,
with an orbital period of 400 yr (top panels) and 800 years (middle panels).  
The bottom panels show the observed $^{12}$CO (2-1) brightness distributions. 
Obviously, the actual gas distribution is more complex than simulated by our 
simple model, but the longer period of 800 years yields images
much more similar to those observed. The shell-intershell density contrast appears 
more clearly on the narrow velocity maps (left panels).

The most likely model, according to our simulations, is that of a strongly enhanced mass-loss rate
triggered every $\simeq 800$ yr by the fly-by of a $\simeq 1\,M_\odot$ companion 
\citep[the mass of cW~Leo
is believed to be $\simeq 2 M_\odot$,][]{Guelin1997} on an orbit almost 
perpendicular to the light of sight. The orbit is probably excentric and the high mass-loss episodes
probably occur near periastron; they last long enough for CW~Leo to change its orbital velocity by 1-2 rd
so that the outer envelope appears asymmetric.

%Figure 12
\begin{figure}
\includegraphics[angle=0,scale=.41]{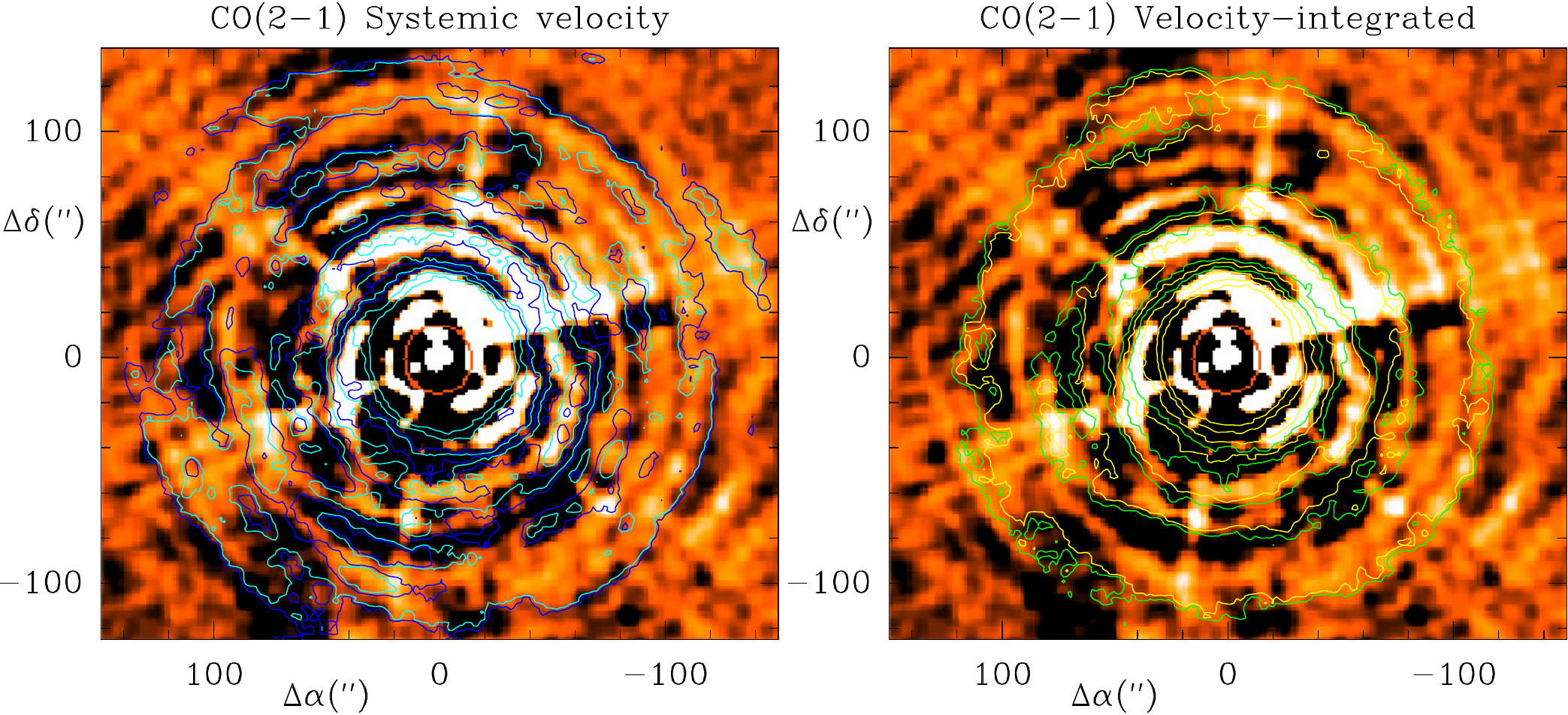} 
\caption{Contour levels of the  CO(2-1) line intensity superimposed on the PSF-deconvolved, 
halo-subtracted 100$\mu$m map of PACS \citep[Fig. 2 of][]{Decin2011}. 
({\it Left}: CO in the velocity interval (V-V$_*$)=$\pm$\,2\,km\,s$^{-1}$; {\it right} CO integrated 
over all velocities). Light blue contours levels: 1 to 6 K by steps of 1 K, yellow contours: 10 
to 70 K.kms$^{-1}$ by 1 K.kms$^{-1}$. The dark blue contour levels represent 0.9 times the adjacent 
light blue contour levels and the green contours 0.85 times the adjacent yellow contours.} 
\label{fig12}
\end{figure}

In this respect, it is instructive to compare both the simulated maps and the
observed CO maps with the map of the dust FIR emission reported by \citet{Decin2011}.
This is done in Fig. 12, where we superimpose on the PSF-deconvolved 
and halo-subtracted PACS map \citep[Fig.2 of][]{Decin2011} the intensity contour levels of our 
$^{12}$CO(2-1) maps (lower panels of Fig. 11). Both the integrated-velocity CO contour levels and 
the CO contours with (V-V$_*$)=$\pm$ 2\,km\,s$^{-1}$ are shown. Because dust and gas should be coeval
and because the dust emission is optically thin at 100$\mu$m, dust emission should be preferably compared 
to the integrated-velocity CO map. Several bright FIR arc-like features appear to closely 
follow the bright rims of the CO shells (e.g. the rim 110$''$ N from the star, the rims 50$''$ 
and 100$''$SW from the star), albeit some dark FIR areas also fall atop bright CO rims (e.g. at 
90$''$ W from the star). Although artefacts caused by the PACS PSF deconvolution make the comparison difficult, 
an overall correlation of the CO and FIR dust emissions seems clear.

Observations of the CO(2-1) emission, at a much higher
resolution, are in progress with the SMA, the PdB Interferometer, and 
for the inner envelope, ALMA. The CO shells,
viewed with an angular resolution of $3''$ or higher appear thinner 
and more clumpy  than in Figure~\ref{fig1}. These differences make
it ineffectual to try to simulate  the pattern of Figure~\ref{fig1},
which is affected by beam smearing, more
precisely. A definitive assessment the
binary scenario must await completion of the interferometer maps.  

After the submission of this article, we learned about the publication of a first 
ALMA study of the close surroundings ($r<3''$) of CW~Leo \citep{Decin2014}, and the discovery of  
a faint point-like object near CW~Leo that could be a companion star \citep{Kim2014}. 
The $^{13}$CO (6-5) emission is resolved by ALMA into a central source plus a couple of arc-like 
features at $\simeq 2''$ from
CW~Leo. Those are interpreted as part of a spiral structure induced by a companion star. The period 
(55 years) and orbit (diameter 20-25 AU, seen edge-on) of such a companion are however 
quite different from those of our model and 
would be difficult to reconcile with our outer envelope observations, in particular with the cusped 
molecular line shapes and the spherical 3D shape of the observed shells. 
The small field of view of the ALMA observations and lack of short spacings (which causes 
a negative lobes in the maps) preclude following the arc-like features at radii larger or smaller than 2$''$,
making the identification of a spiral very tentative for the time being. The arc-like features, on the other hand, 
could be pieces of a new shell, ejected some 80 yr ago when our model companion was near periastron.

The point-like object, discovered by \citet{Kim2014} on 2011 HST images, is interpreted by these authors as 
a possible companion M star. Its {\it apparent} distance from CW~Leo (0.5$''$, or 65 AU) is compatible with that of 
our model companion.     

\section{Conclusion}
Our $^{12}$CO(2-1) and $^{13}$CO(2-1) line emission maps of IRC+10216,
made with the IRAM 30-m telescope, reveal the presence of over-dense
spherical shells, some of which have already been noticed on optical
images or in the dust thermal emission. The CO emission peaks on the
central star, CW~Leo, but remains relatively strong up to $r=180''$
from that star. Its intensity drops or vanishes further
out. We interpret the sudden decrease of the CO brightness to
photodissociation and set the photodissociation limit $r_{phot}$ to
that radius.

The CO envelope fits well inside the large bow-shock discovered by \citet{Sahai2010}
and must be freely expanding into the
cavity cleared up by the shock. Outside the tiny ($<1''$) dust
formation region where the expelled matter is accelerated, the gas
expands radially at a remarkably constant velocity: 14.5
km\,s$^{-1}$. For a distance of 130 pc, the photodissociation radius
corresponds to a look-back time of $\simeq 8000$ yr. Because of CO
self-shielding, CO should be a reliable tracer of the molecular gas up
to $r_{phot}$, and the $^{12}$CO(2-1) and $^{13}$CO(2-1) maps must
teach us about the mass loss history in that period of time.

The over-dense shells show that the mass loss process is highly
variable on timescales of hundreds of years.  Previous studies suggested
that the average mass loss rate has strongly decreased in the last few
thousand years. Comparing the intensities of the CO lines near the
star, across the envelope and beyond $r_{phot}$ we do not see evidence
of such a decrease: the mass-loss rate averaged over periods of
$10^3$ yr appears to be 2-4 10$^{-5}\,
M_\odot$\,yr$^{-1}$ and to stay constant within a factor of 2, 
which is about the accuracy
attached to our mass derivation method.
   
The velocity-channel $^{12}$CO(2-1) emission maps, particularly the
map at the star velocity that  traces the molecular gas in the plane
of the sky, show a succession of bright circles that denote the rims
of the over-dense shells. The circles are not concentric as one may
expect in the case of a spherical outflow, but shifted by several arcsec in different directions; the
outermost shells extend farther to the N-NE. The typical shell
separation is $800-1000$ yr and seems to increase outwards. The
shell-intershell brightness contrast is $\geq 3$ in the outer
envelope.

These key features can all be accounted for if CW~Leo has a companion
star with an excentric orbit and if the mass loss increases when
the companion is close to periastron. Another mass loss mechanism
that has been proposed is a cyclic magnetic activity at the stellar
surface, similar to that on the Sun \citep{Soker2000}.
Mira-type oscillations and thermal pulses seem to be ruled out because their
periods are much shorter or longer than implied by the shell
separation. The angular resolution of the CO observations reported
here does allow us to not properly resolve the shells and their substructures. Higher angular
resolution observations are currently in progress at the SMA,
PdBI, and ALMA. Hopefully, they will enable us to decide on the mass ejection
mechanism and yield more reliable values of the mass-loss rate and of the 
shell-intershell density
contrast. The latter is of great importance for a better understanding
of circumstellar chemistry \citep[see e.g.][]{Guelin1999,Cordiner2009}.  

\begin{acknowledgements}

We thank Spanish MICINN for funding support
through grants AYA2006-14876, AYA2009-07304, and the CONSOLIDER
program "ASTROMOL" CSD2009-00038. 
MG acknowledges 
support from the CNRS program PCMI, as well as from the SMA. 
\end{acknowledgements}

\end{document}